\documentclass{article}

\PassOptionsToPackage{numbers, compress}{natbib}

\usepackage{utfsym}
\usepackage{enumitem}

    \usepackage[final]{neurips_2025}

\usepackage{amsmath}
\usepackage[utf8]{inputenc} 
\usepackage[T1]{fontenc}    
\usepackage{hyperref}       
\usepackage{url}            
\usepackage{booktabs}       
\usepackage{amsfonts}       
\usepackage{nicefrac}       
\usepackage{microtype}      
\usepackage{xcolor}         
\usepackage{graphicx}
\usepackage{multirow}
\usepackage{autobreak}
\usepackage{tablefootnote}
\usepackage{makecell}
\usepackage{wrapfig}
\usepackage{subfigure}
\title{SongBloom: Coherent Song Generation via Interleaved Autoregressive Sketching and Diffusion Refinement}

\author{%
Chenyu Yang\textsuperscript{\textnormal{1,4}}\quad\qquad
Shuai Wang\textsuperscript{\textnormal{3,4,}\thanks{Corresponding authors.}}\quad\qquad
Hangting Chen\textsuperscript{\textnormal{2}}\quad\qquad
Wei Tan\textsuperscript{\textnormal{2}}\\
\textbf{
Jianwei Yu\quad\qquad
Haizhou Li\textsuperscript{\textnormal{1,4,*}}
}
\vspace{6pt}\\
\textsuperscript{1}The Chinese University of Hong Kong, Shenzhen \quad
\textsuperscript{2}Tencent AI Lab  \\
\textsuperscript{3}Nanjing University \quad
\textsuperscript{4}Shenzhen Research Institution of Big Data
}

\begin{document}

\maketitle

\setcounter{footnote}{0}
\begin{abstract}
Generating music with coherent structure, harmonious instrumental and vocal elements remains a significant challenge in song generation. Existing language models and diffusion-based methods often struggle to balance global coherence with local fidelity, resulting in outputs that lack musicality or suffer from incoherent progression and mismatched lyrics. This paper introduces \textbf{SongBloom}\footnote{Code: \url{https://github.com/Cypress-Yang/SongBloom}}, a novel framework for full-length song generation that leverages an interleaved paradigm of autoregressive sketching and diffusion-based refinement. SongBloom employs an autoregressive diffusion model that combines the high fidelity of diffusion models with the scalability of language models. 
Specifically, it gradually extends a musical sketch from short to long and refines the details from coarse to fine-grained. The interleaved generation paradigm effectively integrates prior semantic and acoustic context to guide the generation process.
Experimental results demonstrate that SongBloom outperforms existing methods across both subjective and objective metrics and achieves performance comparable to the state-of-the-art commercial music generation platforms. 
Audio samples are available on our demo page: 
\url{https://cypress-yang.github.io/SongBloom\_demo}.
\end{abstract}

\section{Introduction}




Music, as one of the most expressive forms of human art, serves as a universal language that transcends cultural and linguistic boundaries. Lyric-to-song generation, which involves producing both vocals and accompaniment from given lyrics, requires comprehensive modeling of diverse musical elements, including instrumentation, vocal expressiveness, structural arrangement, and emotional dynamics. These complexities pose significant challenges for existing generative approaches~\cite{briot2020deep,hernandez2022music}. In particular, end-to-end models that directly synthesize music face considerable challenges in maintaining sound quality, as music’s wide frequency range and complex temporal dynamics impose stringent demands on generative fidelity and model capacity.

Advanced approaches to long-form generation typically adopt either a unified non-autoregressive (NAR) architecture~\cite{ning2025diffrhythm} that models the entire process with a diffusion transformer, or an autoregressive (AR) framework~\cite {yuan2025yuescalingopenfoundation, yang2025songeditor} that employs language models (LMs) and separates the generation into semantic and acoustic stages. However, NAR models often struggle to capture precise alignments between phonemes and audio frames, whereas AR approaches typically predict quantized tokens that suffer from low bitrates, resulting in weak musicality and degraded audio quality.

To address these challenges, we propose SongBloom, a novel lyric-to-song generation approach that combines the strengths of both AR and NAR approaches. SongBloom is built upon an autoregressive diffusion architecture~\cite{li2024autoregressive}, which is well-suited for continuous-valued modalities such as music and speech, and has shown strong scalability and robust long-range generation capabilities~\cite{jia2025ditar} comparable to that of audio language models. Given structured lyrics and a 10-second reference audio clip, SongBloom can generate full-length songs with diverse sections up to 150 seconds long.

To further improve the semantic alignment in long-form song generation, we decouple high-level sketch planning from low-level acoustic synthesis, and integrate both within the unified autoregressive framework that supports joint modeling in the style of Chain-of-Thought (CoT) prompting~\cite{wei2022chain}.
Notably, we observe that the entire sketch sequence is not always necessary for predicting each acoustic frame. Instead, the acoustic context itself can provide valuable guidance for shaping subsequent sketch planning. Based on this insight, we partition both semantic and acoustic sequences into patches and generate them in an interleaved manner. This paradigm not only facilitates bidirectional contextual exchange between sketching and refinement stages but also significantly reduces the sequence length needed during acoustic synthesis.

Both subjective and objective evaluations are conducted in our experiments. The assessment employs a comprehensive suite of metrics covering multiple dimensions, including musicality, audio quality, structural coherence, and overall musical aesthetics. The results show that SongBloom consistently outperforms all open-source baselines and surpasses most commercial systems, achieving state-of-the-art performance across several key metrics.

The main contributions of this paper can be concluded as:
\begin{itemize}[leftmargin=15pt]

    \item We propose SongBloom, the first autoregressive diffusion-based model for full-length song generation, which is capable of producing high-quality and expressive songs with coherent structure and rich acoustic details. 

    \item We design a unified framework that combines coarse and fine stages into a single, jointly optimized model, and treats sketch tokens as CoT-like prompts for acoustic generation directly. Additionally, we introduce a novel interleaved generation paradigm that alternates between sketch and acoustic patches, effectively utilizing both semantic and acoustic contexts.
    
    \item Experiments demonstrate that SongBloom surpasses all non-commercial baselines and achieves competitive performance with the state-of-the-art Suno-v4.5 in terms of both subjective and objective metrics. Additionally, our approach achieves a relatively low real-time factor (RTF) compared to other LM-based methods, indicating a favorable trade-off between generation quality and computational efficiency.

\end{itemize}


\begin{table}
    \centering
    \resizebox{\linewidth}{!}{
    \begin{tabular}{l|ccccc}
    \toprule
    System & Architecture & Size & Max Length & Sample Rate (kHz) & Structure \\
    \midrule
    SongGen~\cite{liu2025songgen} & LM & 1.1B & 30s & 16 & \usym{2715} \\
    SongEditor~\cite{yang2025songeditor} & LM + Diff. & 0.7B + 1B & 120s & 44.1 & \usym{1F5F8} \\
    DiffRhythm-base~\cite{ning2025diffrhythm} & Diff. & 1.1B & 95s & 44.1 & \usym{2715} \\
    DiffRhythm-full~\cite{ning2025diffrhythm} & Diff. & 1.1B & 285s & 44.1 & \usym{2715} \\
    YuE~\cite{yuan2025yuescalingopenfoundation} & LM + LM & 7B + 1B & 300s & 44.1 & \usym{1F5F8}  \\
    \midrule
    SongBloom-tiny & LM \& Diff. & 1.3B & 60s & 48 & \usym{1F5F8}\\
    SongBloom-full & LM \& Diff. & 2B & 150s & 48 & \usym{1F5F8} \\
    \bottomrule
    \end{tabular}
    }
    \caption{Comparison of existing song generation models. “+” indicates two cascaded models, while “\&” indicates one unified model with jointly optimized objectives.}
    \label{tab:intro}
\end{table}

\section{Related work}

\subsection{Song Generation}
Song generation models aim to produce both singing voice and accompanying music. Previous studies, such as Melodist~\cite{hong2024text}, SongGen~\cite{liu2025songgen}, and MelodyLM~\cite{li2024accompanied}, have shown promising results in generating sentence-level singing and accompaniment. However, generating long-form song pieces that encompass diverse musical structures, such as verses, choruses, and instrumental-only sections, remains a significant challenge. JukeBox~\cite{dhariwal2020jukebox} was one of the earliest attempts at generating long-form songs, though it suffers from limited genre and timbre control. SongComposer~\cite{ding2024songcomposer} supports long-form lyric-to-melody generation but struggles to preserve certain aspects such as timbre.

Recent research has increasingly focused on long-form and structure-guided song generation. DiffRhythm~\cite{ning2025diffrhythm} introduces a single diffusion model capable of generating songs up to 285 seconds in length while maintaining a low real-time factor.
YuE~\cite{yuan2025yuescalingopenfoundation} employs a coarse-to-fine generation strategy, leveraging two large-scale language models to produce acoustic tokens with improved fidelity and structure.
SongEditor~\cite{yang2025songeditor} proposes a two-stage framework that supports both full song generation and infilling-based editing tailored to specific lyrics, offering greater flexibility for user-driven customization.
In addition, commercial platforms such as Suno\footnote{\url{https://suno.com}} and Udio\footnote{\url{https://udio.com}} have demonstrated promising capabilities in complete song generation. However, due to the lack of publicly available technical details, these systems cannot be thoroughly evaluated or compared.

\subsection{Autoregressive Diffusion Models}
While language models predominantly operate in discrete-valued space, this paradigm might be suboptimal for inherently continuous audio signals. A common workaround involves transforming waveforms into discrete token sequences through vector quantization (VQ)\cite{vasuki2006review,gray1984vector} or residual vector quantization (RVQ)\cite{barnes1996advances}.
Recent work, however, challenges the necessity of discretization. For instance, \citet{li2024autoregressive} demonstrates that discrete tokens may not be essential for autoregressive language models in image generation. Some research has also explored autoregressive diffusion techniques for text~\cite{wu2023ar} and speech~\cite{benita2023diffar} generation.
ARDiT~\cite{liu2024autoregressive} introduces a decoder-only diffusion transformer for zero-shot text-to-speech generation, achieving impressive results without relying on discrete tokens. DiTAR~\cite{jia2025ditar} further simplifies the architecture through a divide-and-conquer strategy, making the model more adaptable for large-scale deployment.

However, the approaches mentioned above remain restricted to the speech domain, and direct migration faces some limitations. First, songs are typically longer than spoken utterances, imposing stricter requirements on contextual consistency across extended durations. Second, the presence of musical accompaniment leads to a higher semantic density of each frame, making generation more difficult.
Recent works~\cite{zhou2024transfusion,sun2024multimodal} introduce a shared transformer architecture capable of handling both discrete and continuous tokens across different modalities. Inspired by this design, we propose a unified autoregressive framework that integrates both sketch prediction and audio synthesis with distinct training objectives.

\section{Preliminary}
\label{sec:preliminary}
\paragraph{Next-Token Prediction}
Given a sequence of tokens $\mathbf{x} = (x_1, x_2, \ldots, x_{t-1})$, the objective is to model the conditional probability of the next token $P(x_t \mid x_1, x_2, \ldots, x_{t-1})$. 
For an entire sequence $(x_1, x_2, \ldots, x_T)$, the joint probability is factorized as:
\begin{align}
P(x_1, x_2, \ldots, x_T|C) = \prod_{t=1}^{T} P(x_t \mid x_1, x_2, \ldots, x_{t-1}; C),    
\end{align}
where $C$ represents the conditions.

\paragraph{Rectified Flow Matching} Rectified flow-matching~\cite{liu2022flow} defines a linear interpolation between a source point $z_0$ and a target point $z_1$ over time $t \in [0, 1]$:
\begin{align}
    z_t &= (1-t)z_0 + tz_1 \\
    \text{d}z_t &= (z_1 - z_0)\text{d}t =v(z_t, t)\text{d}t
\end{align}
where $z_0$ denotes the original data and $z_1 \sim \mathcal{N}(0, 1)$ is a random noise. $v(z_t, t)$ represents the velocity field guiding the transformation from $z_0$ to $z_1$.
The reverse process enables the generation  by inverting this transformation:
\begin{align}
    z_{t-\Delta t} =  z_t - v(z_t, t)\cdot\Delta t
\end{align}
This formulation allows the model to sample from the data distribution by iteratively denoising from the noise distribution.

\section{SongBloom}
\subsection{Task Formulation}
MusicLM~\cite{agostinelli2023musiclm} proposes a two-stage framework that leverages semantic tokens, referred to as \textit{sketch}, extracted from self-supervised learning (SSL) models such as MERT~\cite{li2023mert} as intermediate representations. In the first stage, an LM is employed to generate the sketch sequence from scratch, as its variable length and discrete nature make it well-suited for language modeling approaches, which have demonstrated strong scalability. 
In the second stage, either diffusion or LM subsequently generates \textit{acoustic} tokens in a coarse-to-fine manner. The overall generation process can be formulated as:
\begin{equation}
p(a_{(0:T]}, s_{(0:T]} \mid C) = p_\theta(a_{(0:T]} \mid s_{(0:T]}, C) \cdot p_\phi(s_{(0:T]} \mid C),
\end{equation}
where $s_i$ denotes the sketch token at frame $i$, and $a_i$ denotes the acoustic token. $T$ is the total length of the frame sequence. The parameters $\phi$ and $\theta$ represent the models responsible for generating semantic and acoustic tokens, respectively. This two-stage architecture has been widely adopted in previous long-form music generation approaches such as~\cite{yang2025songeditor, yuan2025yuescalingopenfoundation, lam2024efficient}.

However, the aforementioned formulation faces several limitations: 1) Future sketch tokens contribute little to the prediction of current acoustic tokens due to the strict token-level alignment between the two representations. As noted in \citet{yang2025songeditor,xu2024mucodec}, using chunk-level semantic-to-acoustic reconstruction introduces minimal degradation in fluency, suggesting that generating the complete semantic sequence in advance may be unnecessary. 2) Given the rich and expressive nature of music, prior acoustic latents can provide valuable contextual cues for sketch token generation. However, existing approaches typically condition only on the semantic history, thereby overlooking potentially useful acoustic information.

To address the aforementioned challenges, we propose a novel generation paradigm that interleaves the generation of sketch and acoustic sequences. Both types of tokens are first segmented into fixed-size patches. The modified generation process is formulated as:
{\small
\begin{align}
p(a_{(0:T]}, s_{(0:T]} \mid C) = \prod_{i=0}^{N} p_\theta(s_{(iP:(i+1)P]}|s_{(0:iP]}, a_{(0:iP]},C) \cdot p_\phi(a_{(iP:(i+1)P]}|s_{(0:(i+1)P]}, a_{(0:iP]},C),
\end{align}}
where $P$ denotes the patch size, $N=\lceil T/P \rceil-1$ denotes the number of patches. $\theta$ and $\phi$ correspond to the parameters responsible for generating sketch tokens and acoustic features, respectively. Although a sequential dependency between the two stages still exists, the interleaved generation paradigm enables bidirectional information exchange between the semantic and acoustic representations. The two stages share a subset of model parameters and are jointly optimized within a unified framework, facilitating coherent and high-fidelity song generation.

\begin{figure}
    \centering
    \includegraphics[width=\linewidth]{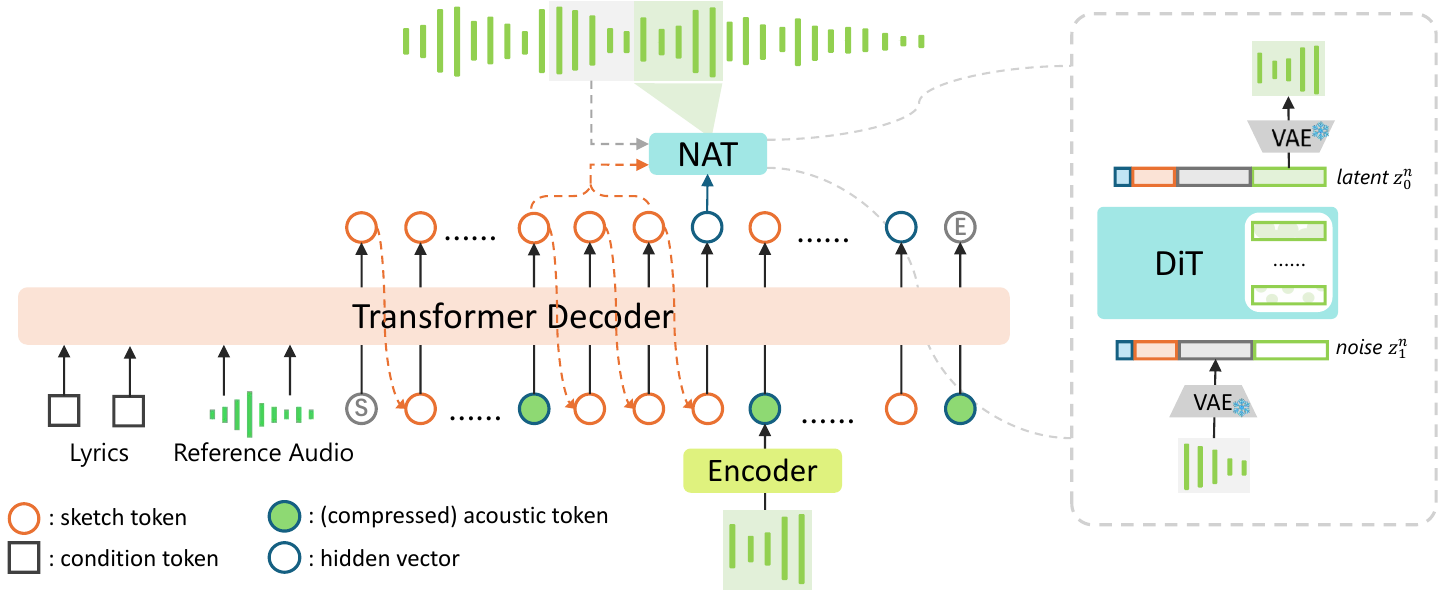}
    \caption{Overall architecture of SongBloom. }
    \label{fig:main}
\end{figure}


\subsection{Data Representation}

\paragraph{Lyric preprocessing}
To incorporate structural information, we introduce two categories of structure flags into the lyric sequence. Vocal-based flags, which indicate the structure of vocal sections (e.g., verse, chorus), are prepended to the beginning of each paragraph. Accompaniment-based flags, which represent non-vocal regions such as intros and outros, are inserted between paragraphs and expanded proportionally according to their exact durations.
All lyric text is normalized and then transformed into a phoneme sequence to serve as input for the subsequent sketch generation stage.

\paragraph{Sketch tokens} In this paper, we adopt the embeddings extracted from MuQ~\cite{zhu2025muq} as our sketch representation. A single vector quantization layer is attached to discretize the sketch embeddings into flattened tokens.  Additionally, we have explored various sketch alternatives including musical signals, which are further discussed in Appendix~\ref{sec:sketch}.

\paragraph{Acoustic latents}
Due to the broader frequency spectrum of music compared to speech, modeling and reconstruction become significantly more challenging. Codec-based discretization methods~\cite{zeghidour2021soundstream, wang2023neural} inherently suffer from a trade-off between reconstruction fidelity and the depth of codebooks, often increasing generation complexity.
To tackle this issue, we substitute the discrete acoustic tokens with continuous latents derived from an autoencoder, which compresses 2-channel 48 kHz music into continuous-valued acoustic latent sequences at a reduced frame rate. This continuous representation preserves high-frequency detail while simplifying the generation process, making it more suitable for high-fidelity music synthesis.

\subsection{Architecture}
\subsubsection{Overview}
The overall model architecture is illustrated in Figure~\ref{fig:main}, which consists of three main modules: an autoregressive transformer decoder, a non-autoregressive diffusion transformer, and an acoustic encoder. Both semantic and acoustic sequences are first segmented into fixed-length patches and then generated in an interleaved manner, enabling the model to maintain consistency between high-level structure and fine-grained acoustic detail.

\subsubsection{Autoregressive Sketch Generation} 
The autoregressive sketch generation stage is designed to generate both sketch tokens $p_\theta(s_{(iP:(i+1)P]}|\cdot)$ and a corresponding hidden vector $p_\theta(h_i|\cdot)$ per patch, which serve as conditioning inputs for the downstream diffusion stage.
The architecture is composed of a stack of transformer decoders with causal masks, enabling left-to-right autoregressive prediction over the sketch tokens. Conditions, including lyric text and style prompt, are prepended to the semantic stream. Each semantic patch is generated based on all previously generated sketch tokens, as well as the acoustic latent features from preceding patches. Specifically, the acoustic features of the current patch are compressed via an acoustic encoder and then inserted as the next token at the position of the hidden vector. This design eliminates the need for custom attention masks, thereby facilitating compatibility with acceleration techniques such as FlashAttention2~\cite{dao2023flashattention2}.

The training objective for the autoregressive sketch generation is formulated as a cross-entropy loss:
\begin{align}
\mathcal{L}_{\text{LM}} = - \frac{1}{NP}\sum_{i=0}^{N-1}\sum_{j=1}^P \log p_\theta(s_{iP+j} \mid s_{<iP+j}, a_{<iP}, C),
\end{align}
where $s_t$ denotes the sketch token at timestep $t$.

\subsubsection{Non-Autoregressive Latent Diffusion}

The non-autoregressive diffusion module predicts acoustic latents within each patch in parallel using a full-attention DiT~\cite{peebles2023scalable} architecture. The model is trained with the Rectified Flow-Matching (RFM) objective~\cite{liu2022flow}, which aims to predict the velocity field $v_\phi(\cdot)$ governing the dynamics of the latent trajectories:
\begin{align}
\frac{\mathrm{d}\textbf{z}^i_{t}}{\mathrm{d}t} = v_\phi(t,\textbf{z}^i_t \mid h_i, s_{(iP:(i+1)P]}, \textbf{z}_0^{i-1}).
\end{align}
Here, $t$ denotes the diffusion time step. In our experiments, we sample $t$ from a logit-normal distribution $\pi_t$ following~\citet{esser2024scaling}. $\textbf{z}^i_t$ represents the latent variables at time $t$ for the $i$-th patch, where $\textbf{z}^i_0 = a_{(iP:(i+1)P]}$ and $\textbf{z}^i_1 \sim \mathcal{N}(\mathbf{0}, \mathbf{I})$ . The variable $h_i$ is the hidden vector inherited from the first stage. $s_{(iP:(i+1)P]}$ refers to the sketch tokens for the current patch, and $\textbf{z}_0^{i-1}$ refers to the acoustic latents from the preceding patch.

The loss function for training is defined as:
\begin{align}
\mathcal{L}_{\text{flow}} = \mathbb{E}_{t\sim \pi_t, \textbf{z}^i} \left[ || v_\phi(t, \textbf{z}^i_t\mid \cdot) - (\textbf{z}^i_1 - \textbf{z}^i_0) ||^2 \right],
\end{align}
where $\pi_t$ denotes the distribution over diffusion steps.
The total training objective combines the sketch generation loss and the diffusion loss:
\begin{align}
    \mathcal{L} = \mathcal{L_{\text{LM}}} + \lambda\cdot \mathcal{L}_{\text{flow}},
\end{align}
where $\lambda=0.1$ is an empirically chosen weighting factor.

Gradients are backpropagated from the diffusion stage to the sketch generation stage via the hidden vector $h_i$. Additionally, the sketch tokens share the same embedding layer across both stages, which improves training efficiency and accelerates convergence.

\citet{jia2025ditar} introduced a classifier-free guidance (CFG) mechanism for autoregressive diffusion models, which is applied only to the hidden vector. In our work, we treat the hidden vector and sketch tokens as a unified conditioning signal, and jointly mask them during training. 
Additional masks are applied solely to the sketch tokens to further encourage the hidden vector to carry more semantic information.

\section{Experimental Setup}

\subsection{Dataset}

We use a large-scale song dataset totaling 100K hours, including both Chinese and English songs, to train our model. The original lyrics are of relatively low quality, prompting the use of a dedicated data cleaning pipeline.  First, we separate the vocal and instrumental tracks using the Demucs model~\cite{rouard2022hybrid}. Then we use the WhisperX toolkit~\cite{bain2022whisperx} to filter out mismatched lyric sentences and recover missing words. The time boundaries of lyrics are also refined based on the results of automatic speech recognition (ASR). Additionally, structural information is extracted from the original tracks using the structure analyzer proposed by \citet{taejun2023allinone}.

For inference, considering that most current song generation models are trained on large-scale but closed-source datasets, it is crucial to ensure that the test samples are not seen during training. To this end, the lyrics are generated using GPT-4o, and the reference audio is randomly clipped from unseen real songs with diverse genres, followed by random shuffling to prevent potential data leakage. A subset comprising 20 samples is used for subsequent human evaluation.

\subsection{Configuration \& Training Setup}
In our experiments, the codebook size for semantic tokens is set to 16384, at a frame rate of 25.
The stable-audio-vae~\cite{evans2025stable} is adopted as the implementation of our waveform autoencoder. Minor modifications are made to its hyperparameters to ensure they have the same frame rate, thereby facilitating synchronized sketches and latents.

The core component of SongBloom is based on the LLaMA-2 decoder architecture~\cite{touvron2023llama}, utilizing causal attention as the backbone of the autoregressive LM. This architecture is further modified to support bidirectional attention, forming our diffusion transformer. Rotary Positional Embeddings (RoPE)~\cite{su2024roformer} are employed in both autoregressive and non-autoregressive transformers to encode positional information. The acoustic encoder is a simple two-layer convolutional network. All conditioning inputs are prepended to the input sequence. Attention modules in each layer use 24 heads with a hidden dimension of 1536, consistent across both autoregressive and non-autoregressive components. The patch size is set to 16, spanning 0.64 seconds. We evaluate two model configurations in our experiments: (1) \textbf{SongBloom-tiny}, comprising 16 layers for the autoregressive LM and 8 layers for the non-autoregressive diffusion transformer, capable of generating songs up to 60 seconds in length; and (2) \textbf{SongBloom-full}, comprising 24 autoregressive layers and 12 non-autoregressive layers, enabling song generation up to 150 seconds.
The former is used for analysis and ablation studies, while the latter is included to enable fair comparisons with other baselines. The model is trained with 16 A100 GPUs for approximately one week.

All models are trained using the AdamW optimizer~\cite{loshchilov2017decoupled} with a learning rate of 1e-4. A cosine learning rate scheduler~\cite{loshchilov2022sgdr} with 2000 warm-up steps is employed to stabilize early training. Each model is trained for approximately 150K steps with a batch size of 128. The DeepSpeed strategy~\cite{rajbhandari2020zero} is adopted to support efficient training.
For inference, both stages share a classifier-free guidance coefficient of 1.5. Next-token prediction is performed using top-k sampling with $k=200$ and a temperature of 0.9. The diffusion process employs the Euler ODE solver with 36 diffusion steps.

\subsection{Evaluation Metrics}

We evaluate the proposed models using the following objective metrics: (1) \textit{Phoneme Error Rate (PER)}, computed based on the separated vocal tracks and the corresponding lyrics; (2) \textit{MuLan Cycle Consistency (MCC)}, which measures the cosine similarity of MuLan~\cite{huang2022mulan} embeddings between generated samples and reference audio or textual descriptions;
(3) \textit{Fréchet Audio Distance (FAD)}\cite{kilgour2018fr}, which quantifies the distributional similarity between generated samples and real songs from which the clips are intercepted;
and (4) \textit{Structural Error Rate (SER)}, which measures the mismatch between detected structural patterns and the target lyric structure. The Dynamic Time Warping (DTW) algorithm~\cite{senin2008dynamic} is first employed to obtain the optimal temporal alignment, and then we calculate the proportion of error duration. Additionally, we leverage the Audiobox-Aesthetic~\cite{tjandra2025aes} to assess musical aesthetics, including content enjoyment (CE), content usefulness (CU), production complexity (PC), and production quality (PQ).

For subjective evaluation, we conduct a Mean Opinion Score (MOS) listening test. A group of at least 10 participants with musical expertise are invited to rate each sample on a scale from 1 to 5. The evaluation focuses on several aspects: musicality (MUS) and audio quality (QLT) assessed separately for both the vocal and accompaniment components, correctness (CRR) of lyrics, and consistency (CST) between samples and prompts.

\section{Results}

\begin{table}[t]
    \centering
    \caption{Objective evaluation results across all models. The upper section reports the performance of commercial platforms (valid until June 23, 2025), while the lower section presents results for open-source baselines as well as our proposed models. }
    \resizebox{\linewidth}{!}{
    \begin{tabular}{l|c|ccccccccc}
    \toprule
      \multirow{2}{*}{Models} & \multirow{2}{*}{Prompt}   &  \multirow{2}{*}{PER(\%)$\downarrow$}  &  \multirow{2}{*}{MCC$\uparrow$} &  \multirow{2}{*}{FAD$\downarrow$} &  \multirow{2}{*}{SER(\%)$\downarrow$} & 
    \multicolumn{4}{c}{Aesthetic Score} & \multirow{2}{*}{RTF$\downarrow$} \\
    \cline{7-10}
       &   & & &  &  & CE$\uparrow$ & CU$\uparrow$ & PC$\uparrow$ & PQ$\uparrow$  & \\
    \hline
     Suno-v4.5~\dag & text & 24.67  & 0.69  & \underline{3.39}  &  \textbf{10.43} & \underline{7.77}  & \underline{7.93} & 6.03 &  8.40    & - \\
     Udio-v1.5  &   10s wav   & 20.04  & 0.79   &  4.04 &  17.92 &  7.47 & 7.63 &  \underline{6.29} &   8.20   & - \\
     Haimian  &   text   & 10.03  & 0.63   & 5.45  &  13.39 & 7.55  &   7.87 & 5.75  &  8.28 & -  \\
     Mureka-O1  &   10s wav   & 7.79  & \underline{0.86}   &  \underline{3.39} & 31.37  & 7.69  & 7.84   & \textbf{6.41}  & \underline{8.45}   & -  \\
    \midrule
    ACE-step~\cite{gong2025ace} &    text  &  54.34 & 0.62  &  8.02 &   \underline{12.27}  & 7.37  & 7.52 &  6.26  &  7.85 & -  \\
     YuE-7B~\cite{yuan2025yuescalingopenfoundation}  &  text + 30s wav    & 27.30  & 0.62  &  5.99 &  13.93 & 7.25 &  7.59 &   5.96   &  8.03 &  13.724 \\
     DiffRhythm-full~\cite{ning2025diffrhythm}\ddag  &   10s wav   & 15.77  & 0.70  & 5.04  & 46.62  &   5.81  & 7.29  &  4.52  & 7.73  & \textbf{0.034}  \\
     SongEditor~\cite{yang2025songeditor}  &  10s wav  &  16.20 & 0.77  &  4.85 &  18.06 & 7.44 & 7.80  &   6.06   &  8.27 & 1.717  \\
     SongBloom-full &    10s wav  &  \underline{6.75} & \textbf{0.88}  &  3.43 &  17.67 & 7.71 &   7.88  &  5.86  &  8.43 & \underline{1.649}  \\
      SongBloom-full-ft &    10s wav  &  \textbf{5.49} & \underline{0.86}  & \textbf{3.20}  & 14.50    & \textbf{7.79}  & \textbf{7.96} &  5.88  &  \textbf{8.47} & \underline{1.649}  \\ 
    \bottomrule
    \end{tabular}
    }
    \footnotesize{\raggedright 
    \dag DiffRhythm requires additional sentence-level timestamps. We first extract paragraph-level timestamps from ASR results of generated samples, then employ GPT-4o to predict the start time of each sentence.
    
    \par
    }

    \label{tab:main}
\end{table}

\begin{table}[t]
    \centering
    \caption{Subjective evaluation results. Consistency (CST) is reported only for models that accept a 10-second reference audio as the style prompt.}
    \resizebox{\linewidth}{!}{
    \begin{tabular}{l|cccccc}
    \toprule
    Models   & MUS$_V$$\uparrow$ & MUS$_A$$\uparrow$ & QLT$_V$$\uparrow$ & QLT$_A$$\uparrow$ & CRR$\uparrow$ & CST$\uparrow$  \\
    \midrule
    Suno-v4.5 & \underline{3.87} $\pm$ 0.28 & \textbf{4.04} $\pm$ 0.18 & 3.83 $\pm$ 0.12 & \textbf{3.96} $\pm$ 0.06 & 2.95 $\pm$ 0.14 & - \\
    Udio-v1.5 & 3.28 $\pm$ 0.27 & 3.55 $\pm$ 0.25 & 3.62 $\pm$ 0.23 & 3.74 $\pm$ 0.21 & 2.57 $\pm$ 0.19 & 2.76 $\pm$ 0.31 \\
    Haimian & 3.39 $\pm$ 0.26 & 3.65 $\pm$ 0.24 & \underline{3.86} $\pm$ 0.17 & 3.90 $\pm$ 0.11 & 3.09 $\pm$ 0.20 & - \\
    Mureka-O1 & \textbf{3.91} $\pm$ 0.26 & \underline{3.93} $\pm$ 0.14 & 3.85 $\pm$ 0.18 & 3.89 $\pm$ 0.12 & \underline{3.38} $\pm$ 0.16 & 3.41 $\pm$ 0.39\\
    \midrule
    ACE-step~\cite{gong2025ace} & 2.95 $\pm$ 0.29 & 2.93 $\pm$ 0.32 & 2.66 $\pm$ 0.27 & 2.68 $\pm$ 0.25 & 2.11 $\pm$ 0.16 & - \\
    YuE-7B~\cite{yuan2025yuescalingopenfoundation}  & 2.93 $\pm$ 0.27 & 3.15 $\pm$ 0.24 & 3.16 $\pm$ 0.27 & 3.28 $\pm$ 0.23 & 2.54 $\pm$ 0.19 & - \\
    DiffRhythm-full~\cite{ning2025diffrhythm} & 2.99 $\pm$ 0.27 & 3.48 $\pm$ 0.27 & 2.97 $\pm$ 0.21 & 3.33 $\pm$ 0.26 & 2.51 $\pm$ 0.23 & 2.45 $\pm$ 0.28 \\
    SongEditor~\cite{yang2025songeditor} & 2.90 $\pm$ 0.32 & 3.11 $\pm$ 0.27 & 3.03 $\pm$ 0.32 & 3.24 $\pm$ 0.28 & 2.89 $\pm$ 0.16 & 2.44 $\pm$ 0.28\\
    SongBloom-full & 3.59 $\pm$ 0.25 & 3.60 $\pm$ 0.23 & 3.83 $\pm$ 0.09 & 3.81 $\pm$ 0.08 & 3.27 $\pm$ 0.17 & \textbf{3.62} $\pm$ 0.27 \\
    SongBloom-full-ft & \textbf{3.91} $\pm$ 0.24 & 3.92 $\pm$ 0.12 & \textbf{3.95} $\pm$ 0.04 & \underline{3.93} $\pm$ 0.10 & \textbf{3.42} $\pm$ 0.18 & \underline{3.45} $\pm$ 0.31\\
    \bottomrule
    \end{tabular}
    }
    \label{tab:subject}
\end{table}



\subsection{Full-Length Lyric-to-Song Generation}
Based on the SongBloom-full model, we fine-tuned it for 1,000 additional steps on synthesized data with a clear alternating structure of verses and choruses, denoted as \textbf{SongBloom-full-ft}. This fine-tuning process enhances the model’s ability to emulate synthesized lyric compositions, leading to improved performance on lyrics with similar stylistic patterns.

\paragraph{Comparison of objective metrics}
As shown in Table~\ref{tab:main}, SongBloom demonstrates superior performance across various metrics and is competitive with Suno-v4.5, the state-of-the-art commercial song generation platform. After fine-tuning on downstream data, SongBloom-full-ft even outperforms Suno-v4.5 in several metrics.

During our evaluation, we observed that Suno tends to follow rigid structural patterns, such as redundantly repeating the chorus at the end, which leads to structural hallucinations and degraded PER performance. In contrast, SongBloom adheres more faithfully to the structure of the input lyrics, enabling flexible and structurally coherent song generation that significantly reduces the PER. 
In terms of the MCC metric, waveform-based style prompts generally possess a natural advantage over text descriptions, as they always provide a comprehensive and unbiased representation of all musical components. Among these approaches, SongBloom-full achieves the highest MCC score.
In terms of the automated-evaluated aesthetic scores, SongBloom-full-ft outperforms all other baselines in three out of the four metrics, further demonstrating its ability to generate high-quality, musically coherent, and aesthetically pleasing songs.

Apart from generation performance, we also assessed the inference speed of SongBloom compared to other open-source models. The integrated design of SongBloom enables outstanding inference efficiency, yielding a lower RTF than all other autoregressive baselines. 
Compared to SongEditor, our model achieves a similar RTF despite having a larger size, since SongEditor takes the entire semantic sequence as input during the diffusion stage, leading to unnecessary computational overhead. Meanwhile, YuE utilizes two LMs and generates multi-layer codec tokens in a flattened pattern, significantly extending the generation sequence. While autoregressive models are naturally slower than their non-autoregressive counterparts, SongBloom achieves an excellent trade-off between efficiency and performance.


\paragraph{Comparison of subjective metrics}
Table~\ref{tab:subject} shows that SongBloom and Suno-v4.5 dominate the human evaluation results, outperforming all other models by a clear margin. Suno-v4.5 demonstrates particular strength in accompaniment generation, whereas SongBloom excels in metrics related to vocal tracks. Notably, SongBloom achieves the highest correctness score, indicating a stronger adherence to the provided lyrics compared to other systems. This suggests that its interleaved generation paradigm effectively preserves semantic intent throughout the song. Furthermore, SongBloom’s superior performance in consistency is corroborated by its high MCC score, reflecting its ability to maintain thematic coherence over every section of songs.

\subsection{Ablation Study on Diffusion Conditions}

Table~\ref{tab:ablation} compares the performance of SongBloom-tiny under different combinations of diffusion conditions. In the absence of sketch tokens, the LM stage produces only a sequence of hidden vectors, and the diffusion stage reconstructs the audio solely based on them. As shown in the table, the sketch plays a critical role in generation. Without it as a coarse-grained CoT, the model fails to learn proper alignment between phoneme sequences and audio, resulting in extremely high PER. In contrast, when the sketch is provided, even without additional conditions, the diffusion transformer can generate intelligible vocals solely based on the hidden vector from the LM stage. Meanwhile, incorporating both the acoustic context and sketch tokens as input conditions during the diffusion stage further enhances both objective and subjective performance, leading to higher-quality generation results.

\begin{table}[t]
    \centering
    \caption{Ablation study of SongBloom-tiny under different diffusion conditions. "H" represents the hidden vector, "C" represents the acoustic context from the previous patch, and "S" represents the sketch tokens of the current patch.}
    \resizebox{\linewidth}{!}{
    \begin{tabular}{clccccccccccc}
    \toprule
    \multirow{2}{*}{w/ sketch} & \multirow{2}{*}{Conditions} && \multicolumn{2}{c}{Objective} && \multicolumn{4}{c}{Aesthetic Score} && \multicolumn{2}{c}{Subjective} \\
    \cline{4-5}\cline{7-10} \cline{12-13}
     & && PER(\%)$\downarrow$ & FAD$\downarrow$ && CE$\uparrow$ & CU$\uparrow$ & PC$\uparrow$ & PQ$\uparrow$ && MUS$\uparrow$ & QLT$\uparrow$ \\
    \midrule 
    \multirow{3}{*}{\usym{1F5F8}} & H+C+S  && \textbf{9.44} & 5.60 && \textbf{7.57} & \textbf{7.66} & 6.00 & \textbf{8.25}  &&  \textbf{3.55} & \textbf{3.77} \\
    & H+C  && 10.49 & \textbf{5.16} && 7.43 & 7.57 & \textbf{6.42} & 8.15  &&  3.43 & 3.46 \\
    & H && 11.35 & 7.58 && 7.46 & 7.50 & 5.80 & 8.09  &&  3.14 & 3.41 \\
    \midrule
    \usym{2715} & H+C && 109.76 & 8.86  && 7.06 & 7.40 & 5.92 & 7.71 && - & - \\
    \bottomrule
    \end{tabular}
    }
    \label{tab:ablation}
\end{table}

\subsection{Effect of Hyper-Parameters}

\begin{figure}
  \centering
  \begin{minipage}[t]{0.53\textwidth}
    \centering
    \includegraphics[width=\textwidth]{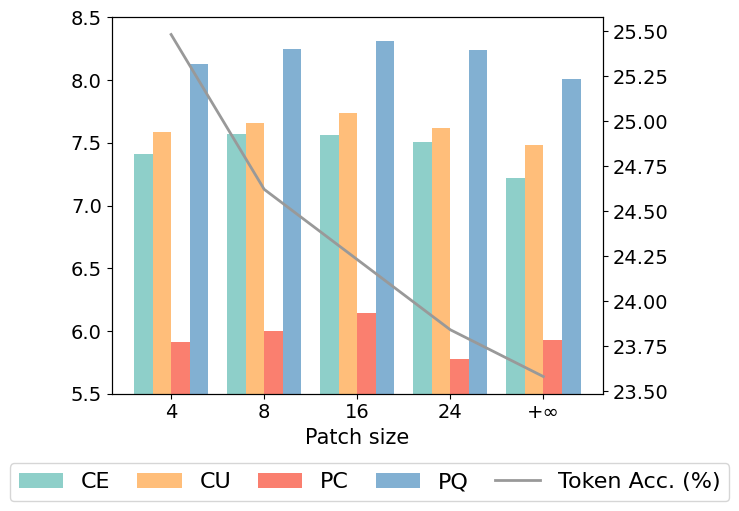}
    \caption{Aesthetic scores and sketch token accuracy of SongBloom-tiny with various patch sizes.}
    \label{fig:patch}
  \end{minipage}
  \hfill
  \begin{minipage}[t]{0.45\textwidth}
    \centering
        \includegraphics[width=\textwidth]{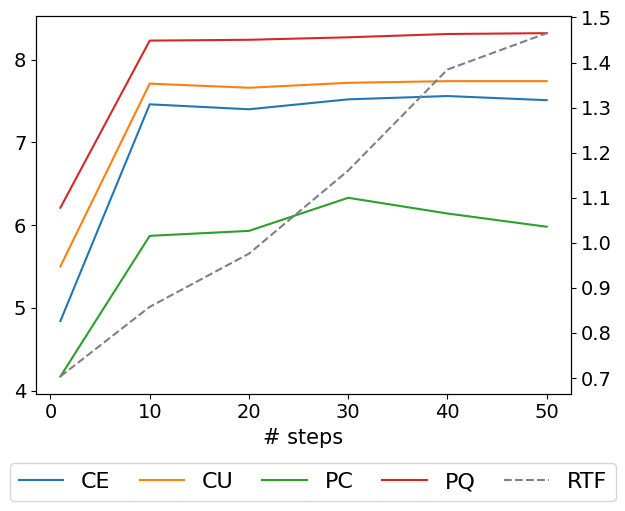}
    \caption{Aesthetic scores of SongBloom-tiny with increasing diffusion steps.}
    \label{fig:steps}
  \end{minipage}
\end{figure}

Figure~\ref{fig:patch} illustrates the impact of various patch sizes, where "$+\infty$" denotes separating the generation of sketch and acoustic sequence. For speech-oriented autoregressive diffusion models~\cite{jia2025ditar, liu2024autoregressive}, a small patch size is essential, as larger patch sizes make the model difficult to converge. For SongBloom, since the sketch tokens have served as the CoT of hidden vectors and guide the diffusion directly, a small patch size is no longer the optimal choice. Although smaller patch sizes provide more acoustic information for sketch generation, thereby improving sketch token accuracy, they also hinder the fluency of acoustic latents during the diffusion stage, as the window for preceding contexts becomes smaller. 
Figure~\ref{fig:steps} illustrates the relationship between performance and diffusion steps during inference. The RTF increases proportionally with the number of diffusion steps, while near-optimal generation performance is achieved in as few as 10 steps, indicating that the inference process can be further accelerated.

\section{Discussions and Limitations}
In this paper, we propose SongBloom, a novel approach for full-song generation that produces expressive, high-quality songs and achieves state-of-the-art performance across multiple metrics. Nevertheless, several open challenges remain. First, the current sketch representation is derived from SSL models, which lack interpretability. Replacing these with some symbolic formats could enable more fine-grained control and user customization, which will be our future work. Second, we believe that some reinforcement learning-based techniques, such as DPO~\cite{rafailov2023direct} or PPO~\cite{schulman2017proximal}, can be applied to SongBloom, which aligns the generation process with user preferences, thereby enabling outputs that better match human aesthetic judgments.

We fully acknowledge the potential ethical risks associated with music generation models. We ensure that both our models and training data are strictly used for academic research purposes only. We respect the intellectual property rights of all original artists and content creators. Every effort has been made to avoid the use of copyrighted material without proper authorization.

\section{Acknowledgement}

This work is supported by National Natural Science Foundation of China (Grant No. 62271432), Shenzhen Science and Technology Program (Shenzhen Key Laboratory, Grant No. ZDSYS20230626091302006), Program for Guangdong Introducing Innovative and Entrepreneurial Teams, Grant No. 2023ZT10X044.

\bibliographystyle{IEEEtranN}
\bibliography{references}
\appendix
\newpage

\section{Ablation Study on Different Sketch Choices}\label{sec:sketch}
Table~\ref{tab:sketch_type} demonstrates our early efforts at decomposing the semantic information of sketches.
We compare different sketch representations, including pitch, pitch + chromagram, and semantic embeddings. As shown in the table, simple pitch-based sketches offer limited alignment capability, resulting in high PER and suboptimal performance in other metrics. The inclusion of chroma features slightly improves alignment, but still falls short of fully capturing the rich semantics needed for coherent vocal generation. 
These results validate the importance of incorporating abstract, semantically meaningful information into the sketch stage, and lay the groundwork for future exploration of interpretable yet powerful sketch formats.

\begin{table}[h]
    \centering
    \begin{tabular}{l|cc|ccccc}
    \toprule
    Type & PER(\%)$\downarrow$ &  FAD$\downarrow$ & CE$\uparrow$ & CU$\uparrow$ & PC$\uparrow$ & PQ$\uparrow$ \\
    \midrule
    no sketch & 109.76 & 8.86  & 7.06 & 7.40 & 5.92 & 7.71\\
    pitch & 103.22 & 5.43 &  7.34 & 7.58 & 6.03 & 8.04 \\
    pitch + chroma & 68.47 & 5.47 & 7.37  & 7.57 & 5.98 & 8.06 \\
    SSL embedding & 9.44 & 5.60 & 7.55 & 7.66 & 6.00 & 8.25 & \\
    \bottomrule
    \end{tabular}
    \caption{Impact of sketch types on SongBloom’s performance.}
    \label{tab:sketch_type}
\end{table}

\section{Time Complexity of SongBloom Inference}
We analyze the inference time complexity of SongBloom compared to a decoupled two-stage model, assuming both have the same number of layers.

Let $L_1$ denote the number of layers in the language model stage, and $L_2$ the number of layers in the diffusion stage. Let $T$ be the total number of frames for both semantic and acoustic sequences (eg. $30~\text{s} ~\times 25~\text{fps} =600$), $P$ the patch size, $N = T / P$ the number of patches, and $S$ the number of diffusion steps.

Assuming key-value caching is used during inference, we analyze the leading-order time complexity:

\textbf{Decoupled 2-stage model:}
\begin{align*}
    O_0 = L_1 \cdot \frac{(1+T)T}{2} + L_2 \cdot (2T)^2 \cdot S
\end{align*}

\textbf{SongBloom:}
\begin{align*}
    O_1 = L_1 \cdot \frac{(1 + T + N)(T + N)}{2} + L_2 \cdot (2P)^2 \cdot N \cdot S
\end{align*}

Substituting $N = T / P$, we compute the difference:
\begin{align*}
    O_1 - O_0 
    &= \left( \frac{L_1}{2P^2} + \frac{L_1}{P} - 4L_2S \right) T^2 
    + \left( \frac{L_1}{2P} + 4L_2PS \right) T
\end{align*}

When $T$ is sufficiently large, the $T^2$ term dominates. In most practical cases, where:
\[
L_1 < 4SP \cdot \frac{2P}{2P + 1}\cdot L_2
\]
the coefficient of $T^2$ is negative. Therefore, we conclude that SongBloom is asymptotically more efficient than the decoupled two-stage models, owing to its patch-wise diffusion mechanism and reduced per-step input length during inference.

\section{Criteria of the Subjective Listening Test}

\begin{enumerate}
    \item \textbf{Musicality of vocal}: (1–5 points) Does the main melody of the generated vocal match the subjective expectation? 
    \begin{itemize}
    \item \textbf{5 points}: The melody is pleasant and emotionally expressive, with strong musical phrasing. It aligns well with expectations.
    \item \textbf{4 points}: The melody generally meets expectations and conveys the song's theme and emotion, but lacks standout features.
    \item \textbf{3 points}: The melody mostly aligns with expectations and conveys the theme and emotion, though some notes feel abrupt.
    \item \textbf{2 points}: Only parts of the melody are coherent; most notes are scattered, and the theme and emotion are vaguely presented.
    \item \textbf{1 point}: The melody significantly deviates from expectations, lacks coherent musical phrasing, and fails to convey the song's theme and emotion.
    \end{itemize}
    \item \textbf{Musicality of accompaniment}: (1–5 points) Does the accompaniment of the generated song sound harmonious? 
    \begin{itemize}
        \item \textbf{5 points}: The accompaniment is richly colored and features diverse instrumentation. The melody is beautiful and complements the main melody harmoniously.
        \item \textbf{4 points}: The accompaniment supports the main melody, but uses limited instrumentation or has a generally average melodic performance.
        \item \textbf{3 points}: The accompaniment mostly supports the main melody, with only minor discord. However, it sometimes clashes with the main melody and lacks variety and color in instrumentation.
        \item \textbf{2 points}: Some segments show disorganized instrumentation and monotonous melody, barely supporting the main melody.
        \item \textbf{1 point}: The instrumentation is chaotic and the melody is discordant. There is a clear conflict with the main melody, failing to provide support.
    \end{itemize}
    \item \textbf{Quality of vocal}: (1–5 points) Is the vocal in the generated music clear and bright, with a full high-frequency range? Are there any noises or distortions present?
    \begin{itemize}
        \item \textbf{5 points}: The vocal quality is rich and clear, with no noise, approaching studio-recording quality.
        \item \textbf{4 points}: The vocal quality is relatively clear, with slight noise that is either imperceptible or barely noticeable.
        \item \textbf{3 points}: The vocal quality contains some noise or distortion, but it does not significantly affect the listening experience.
        \item \textbf{2 points}: The vocal quality is unclear and unstable, resulting in a poor listening experience. Noticeable noise or distortion is present.
        \item \textbf{1 point}: The vocal quality is extremely poor, with an unpleasant listening experience, and the vocal characteristics are barely recognizable.
    \end{itemize}
    \item \textbf{Quality of accompaniment}: (1–5 points) Is the high-frequency range of the generated music’s accompaniment full? Are there any noises or instrumental distortions?
    \begin{itemize}
        \item \textbf{5 points}: The accompaniment has a full and clear sound quality with no flaws. The characteristics and melodies of different instruments are clearly distinguishable.
        \item \textbf{4 points}: The accompaniment has good sound quality with slight noise. Only a few instruments in certain segments are hard to distinguish or slightly distorted, but this does not affect the overall listening experience.
        \item \textbf{3 points}: The accompaniment has average sound quality. Some instruments are unclear or unidentifiable in certain segments. There is noticeable noise, distortion, or a lack of clarity.
        \item \textbf{2 points}: The accompaniment has poor sound quality. In most parts of the piece, most instruments are unrecognizable. There is clear noise, distortion, or lack of clarity.
        \item \textbf{1 point}: The accompaniment has extremely poor sound quality, with severe distortion, making it nearly impossible to identify any instrumental characteristics.
    \end{itemize}
    \item \textbf{Correctness of lyrics}: (1–4 points) Does the song content match the lyrics? Are there any errors such as extra words, missing words, or mechanical repetition?
    \begin{itemize}
        \item \textbf{4 points}: The song content fully matches the lyrics, with no missing or extra words, and no mechanical repetition of musical segments.
        \item \textbf{3 points}: The generated song contains a small number (within 5 words) of unclear, repeated, or missing lyrics.
        \item \textbf{2 points}: The generated song contains multiple segments with unclear, repeated, or missing lyrics.
        \item \textbf{1 point}: The generated song does not match the lyrics at all.
    \end{itemize}
    \item \textbf{Consistency of prompt}: (1–5 points) Does the musical style of the generated song match the style of the reference audio prompt?
    \begin{itemize}
        \item \textbf{5 points}: The musical style of the generated song fully matches the style specified in the prompt.
        \item \textbf{4 points}: The musical style of the generated song is similar to the specified prompt, with only slight differences in some segments.
        \item \textbf{3 points}: The musical style is somewhat similar to the specified style, but only vaguely reflects its characteristics.
        \item \textbf{2 points}: The musical style does not resemble the specified style, with only faint traces of the intended musical elements.
        \item \textbf{1 point}: The musical style has no relation to the specified style, making it difficult to connect the prompt with the resulting music.
    \end{itemize}
\end{enumerate}


\newpage

\end{document}